\documentclass[
reprint,
longbibliography,
aps,
prl,
twoside,
superscriptaddress,
bibnotes,
floatfix
]{revtex4-2}

\usepackage{physics}
\usepackage{dsfont}
\usepackage{color,newfloat}
\usepackage{graphicx}
\usepackage{dcolumn}
\usepackage{bm}
\usepackage{amsmath,amssymb,amsthm,amsfonts}
\usepackage{mathtools}
\usepackage{tikz}
\usetikzlibrary{decorations.pathreplacing}
\usepackage{multirow}
\usepackage{hyperref}
\usepackage[all]{hypcap}
\usepackage{chngcntr}
\usepackage{enumerate}
\usepackage[none]{hyphenat}
\usepackage[euler]{textgreek}
\usepackage{xcolor}
\usepackage{tabularx}
\usepackage{makecell}
\newcolumntype{Y}[1]{>{\centering\arraybackslash\hsize=#1\hsize}X}

\begin{document}
\title{Exchange Symmetry in Multiphoton Quantum Interference}
\author{Shreya Kumar}
\thanks{These authors contributed equally.}
\affiliation{Institute for Functional Matter and Quantum Technologies, University of Stuttgart, 70569 Stuttgart, Germany}
\affiliation{Center for Integrated Quantum Science and Technology (IQST), University of Stuttgart, 70569 Stuttgart, Germany}
\author{Alex E Jones}
\thanks{These authors contributed equally.}
\affiliation{Quantum Engineering Technology Labs, H. H. Wills Physics Laboratory and Department of Electrical and Electronic Engineering, University of Bristol, Bristol BS8 1FD, UK}
\author{Daniel Bhatti}
\affiliation{Networked Quantum Devices Unit, Okinawa Institute of Science and Technology Graduate University, Okinawa, Japan}
\author{Stefanie Barz}
\email{barz@fmq.uni-stuttgart.de}
\affiliation{Institute for Functional Matter and Quantum Technologies, University of Stuttgart, 70569 Stuttgart, Germany}
\affiliation{Center for Integrated Quantum Science and Technology (IQST), University of Stuttgart, 70569 Stuttgart, Germany}

\begin{abstract}

Photons are bosons, and yet, when prepared in specific entangled states, they can exhibit non-bosonic behaviour. While this phenomenon has so far been studied in two-photon systems, exchange symmetries and interference effects in multi-photon scenarios remain largely unexplored. In this work, we show that multi-photon states uncover a rich landscape of exchange symmetries. With three photons already, multiple pairwise combinations are possible, where each pair of photons can exhibit either bosonic, fermionic, or anyonic exchange symmetry. This gives rise to \textit{mixed} symmetry systems that are not possible to achieve with two photon alone. We experimentally investigate how these symmetry configurations manifest themselves in the observed interference of three photons. We show that multi-photon interference can be effectively turned on and off by tuning the symmetry of the constituent pairs. The possibility of accessing and tuning new quantum statistics in a scalable photonic platform not only deepens our understanding of quantum systems, but is also highly relevant for quantum technologies that rely on quantum interference. 

\end{abstract}
\maketitle

\begingroup
  \renewcommand{\thefootnote}{\fnsymbol{footnote}}
  \setcounter{footnote}{1}%
  \footnotetext{These authors contributed equally.}
\endgroup

\section{Introduction}
Particles in nature are either bosons or fermions and have wavefunctions that are symmetric or anti-symmetric, respectively, under the exchange of two identical particles \cite{messiah1964,leinaas1977,girardeau1965}. This exchange symmetry directly influences quantum interference: identical bosons bunch together, while identical fermions, constrained by the Pauli exclusion principle, exhibit anti-bunching~\cite{hong1987,davis1995,shchesnovich2016,liu1998}. Nevertheless, theoretical models of particles with more exotic exchange statistics---such as those of anyons, and Majorana fermions---provide foundational insights in topics like quantum error correction and quantum computing~\cite{kitaev2003,dauphinais2017,kesselring2024}.

The overall symmetry of the wavefunction of bosons or fermions must be preserved. However, given access to multiple degrees of freedom, this symmetry may be distributed in a manner that emulates dynamics distinct from those of the fundamental particle symmetry. This means that, for example, with the right preparation, bosonic particles can manifest fermionic or anyonic statistics~\cite{matthews2013,descamps2025}.

Photons are a particularly appealing testbed for exploring particle symmetries in this way. First, they do not interact, and so, in contrast to charged particles like ions, their dynamics are governed solely by symmetries and interference. Second, they possess multiple distinct degrees of freedom---such as polarisation and arrival time---that are relatively straightforward to prepare, control, and measure~\cite{crespi2011,brecht2015,pilnyak2019}. Third, recent technological advances have enabled the preparation of high-quality entangled states of photons, which is necessary to distribute the bosonic symmetry in a way that unlocks other particle statistics~\cite{schaeff2015,thomas2022}. For instance, polarisation-entangled photon pairs with different symmetries in their spatial modes have been used to explore particle statistics going beyond those of bosons~\cite{liu2022}.
This is important, for example, in the context of quantum interference or quantum walks~\cite{van2012,sansoni2012,matthews2013,vetlugin2022}.

Scaling to the interference of larger numbers of photons leads to rich interference phenomena~\cite{lim2005,tichy2012, spagnolo2013,tillmann2015,tichy2017,rodari2024,faleo2024, spivak2022,seron2023}. Recent studies have investigated the roles of properties such as photon distinguishability and state purity in interference~\cite{tichy2011,menssen2017,munzberg2021,minke2021,jones2023}, and multiphoton interference can be harnessed to generate entanglement~\cite{lu2007,pilnyak2017,zhong2018,kumar2023,bhatti2023,bhatti2025}. Furthermore, various quantum technologies, ranging from quantum communication and metrology to quantum simulation, benefit from multiphoton interference~\cite{walther2005,birrittella2012,llewellyn2020}.

In this work, we investigate how exchange symmetry influences the dynamics of three particles. Having states with more than two particles enables different pairings, and consequently, various combinations of pairwise symmetries. We experimentally probe three-photon interference involving quantum states with \emph{mixed}-exchange symmetries---multipartite states in which some components are symmetric under exchange, while others are anti-symmetric. By varying the symmetry of the constituent pairs of photons, we can tune the contribution of the purely bosonic components to the interference. We observe that signatures of multiphoton interference appear only when both pairings are symmetric and disappear in the case of mixed symmetries. Our work shows that photons are an ideal platform for exploring fundamental particle statistics and underlines how the symmetry of states can influence quantum interference.
\begin{figure*}
    \centering
    \includegraphics[width=1\textwidth]{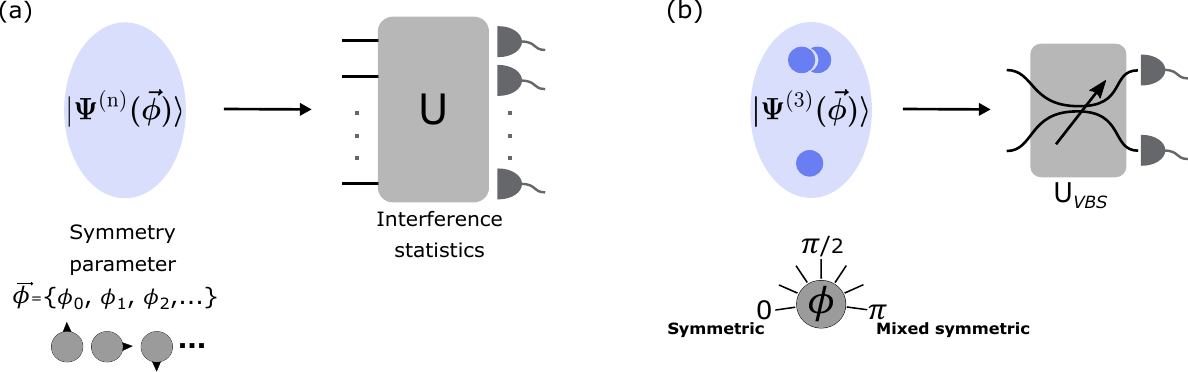}
    \caption{\textbf{Probing exchange symmetry in photon interference:} (a) The bosonic symmetry of the wavefunction, $\ket*{\Psi^{(n)}(\vec{\phi}}$, for $N$ photons can be distributed over different degrees of freedom using a set of parameters $\vec{\phi}$. This allows access to non-bosonic pairwise exchange statistics when the state is evolved under some unitary $U$ and then measured. (b) A state of three photons, $\ket*{\Psi^{(3)}(\vec{\phi}}$, with a relative phase, $\phi$, to directly adjust the exchange symmetry and probe different particle statistics using a variable beam splitter described by the unitary $U_{VBS}$ (see Eq.\ref{eq:VBS}).}
    \label{fig:concept}
\end{figure*}
\begin{table*}
    \centering
    \includegraphics[width=\textwidth]{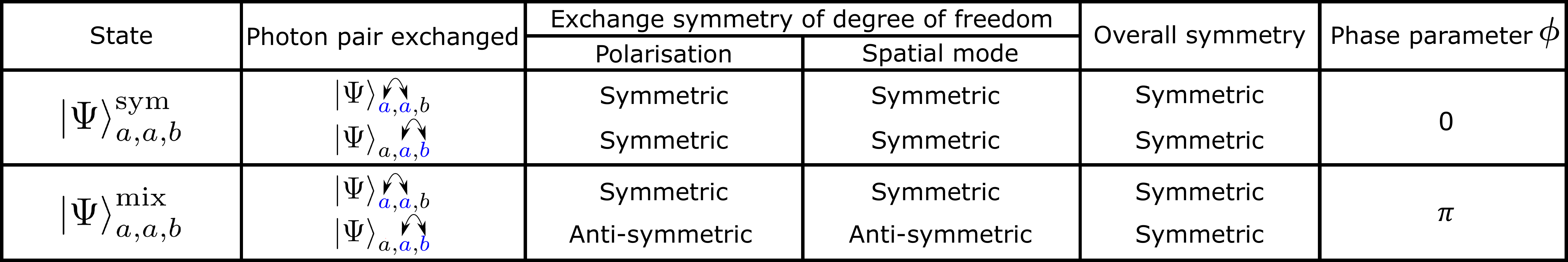}
    \caption{Exchange symmetry of the different degrees of freedom for the symmetric state $\ket{\Psi}_{a,a,b}^{\text{sym}} = \frac{1}{\sqrt{3}}(a_{0}^{\dagger}a_{0}^{\dagger}b_{1}^{\dagger}+a_{0}^{\dagger}a_{1}^{\dagger}b_{0}^{\dagger})\ket{\text{vac}}$, and the mixed-symmetry state $\ket{\Psi}_{a,a,b}^{\text{mix}}=\frac{1}{\sqrt{3}}(a_{0}^{\dagger}a_{0}^{\dagger}b_{1}^{\dagger}-a_{0}^{\dagger}a_{1}^{\dagger}b_{0}^{\dagger})\ket{\text{vac}}$, corresponding to the symmetry parameter $\phi=0$ and $\phi=\pi$, respectively. The indices $a,a,b$ indicate two photons in the spatial mode $a$, and a third photon in mode $b$. Note that a symmetric exchange leaves the state unchanged, whilst an anti-symmetric exchange introduces a global factor of $-1$. However, the overall bosonic symmetry of the state is preserved in all cases.} 
    \label{table:sym}
\end{table*}

\section{Theory}
Photonic states can be described in the context of interference experiments by their \textit{internal} states, i.e., the degrees of freedom that are not resolved by the detection scheme, and \textit{external} states, the degrees of freedom that are resolved~\cite{englbrecht2024}. In this work, we use two internal states, $\ket{0}$ and $\ket{1}$, and two external states, $\ket{a}$ and $\ket{b}$. Our goal is to describe the exchange symmetry of photonic states that contain internal and external degrees of freedom.
For two photons, a possible basis for describing an arbitrary internal state is given by the Bell basis~\cite{sych2009}:
\begin{align}
    \ket{\psi^{\pm}}_{int}&= \frac{1}{\sqrt{2}}(\ket{01}\pm\ket{10}),\\
   \ket{\phi^{\pm}}_{int}&= \frac{1}{\sqrt{2}}(\ket{00}\pm\ket{11}),
\end{align}
This case is well known in the context of quantum information: the Bell basis is complete and the discrimination of Bell states is of central importance in photonic quantum information~\cite{michler1996, kwiat1998, schuck2006, bayerbach2023, hauser2025}.
For the external degree of freedom, let us consider two photons that occupy two spatial modes $a$ and $b$. The corresponding basis for pairs of external states are given by:
\begin{align}
\ket{\psi^{\pm}}_{ext}&= \frac{1}{\sqrt{2}}(\ket{ab}\pm\ket{ba}),\label{eq:ext psi}\\
\ket{\phi^{\pm}}_{ext}&= \frac{1}{\sqrt{2}}(\ket{aa}\pm\ket{bb}).
\end{align}
In this work, we impose the condition that the two photons occupy distinct spatial modes, which makes their exchange symmetry easily accessible, leaving us with the external states $\ket{\psi^{\pm}}_{ext}$ as in Eq.~\ref{eq:ext psi}.
Note that for the internal and external degrees of freedom, the state $\ket{\psi^{-}}_{int/ext}$ is anti-symmetric under pairwise exchange, whereas all others are symmetric. Additionally, the state $\ket{\psi^{-}}_{ext}$ is an eigenstate of the transformation of a beam splitter and, therefore, remains unchanged upon passing through it. 

Photons are bosons, and as such their overall wavefunction exhibits symmetric behaviour. This results in four possible combinations of internal and external states, given by: 
\begin{align}
\ket{\Phi^{+}}&\sim\ket{\phi^{+}}_{int}\ket{\psi^{+}}_{ext}, \\
\ket{\Phi^{-}}&\sim\ket{\phi^{-}}_{int}\ket{\psi^{+}}_{ext}, \\
\ket{\Psi^{+}}&\sim\ket{\psi^{+}}_{int}\ket{\psi^{+}}_{ext}, \\
\ket{\Psi^{-}}&\sim\ket{\psi^{-}}_{int}\ket{\psi^{-}}_{ext},
\end{align}
which are the four canonical Bell states~\cite{zeilinger1998}. We can also interpret this as the distribution of the exchange symmetry over the internal and external degrees of freedom, respectively. The first three states are expressed as the product of states that are symmetric over the internal and external degrees of freedom. The overall bosonic symmetry of the last state $\ket{\Psi^{-}}$ factorises into states that are anti-symmetric over the internal and external degrees of freedom. 


Extending to more than two photons allows for interesting combinations of pairwise symmetries among them, leading to a richer interference landscape and enabling finer control over the symmetry properties of the resulting states. For $N$ photons, the overall state must exhibit bosonic exchange symmetry, and the exchange symmetry can again be distributed over the internal and external degrees of freedom. We can define a symmetric $N$-photon wavefunction $\ket*{\Psi^{(N)}(\vec{\phi})}$ that describes the overall state of $N$ photons, and a set of parameters $\vec{\phi}$, that captures how the exchange symmetry is partitioned across the internal and external states (see Fig.~\ref{fig:concept}(a)). In general, we can decompose the symmetric wavefunction into a superposition of states that are overall symmetric, but possess different symmetry types for constituent degrees of freedom. That is, $\ket*{\Psi^{(N)}(\vec{\phi})}=\sum C_{\mu\mu'}\ket{\psi^{\mu}}_{int}\ket*{\psi^{\mu'}}_{ext}$, where $C_{\mu\mu'}$ is the weight of specific combinations of internal and external states, $\ket{\psi^{\mu}}_{int}$ and $\ket*{\psi^{\mu'}}_{ext}$, respectively~\cite{harrow2005,rowe2012,stanisic2018}.

We aim to study such symmetries with a state of three photons occupying two polarisation (internal) and two spatial (external) modes. Let us define a set of states with specific pairwise exchange symmetry:
\begin{align}
    \ket{\Psi}_{a,a,b}^{\text{sym}}&= \frac{1}{\sqrt{3}}(a_{0}^{\dagger}a_{0}^{\dagger}b_{1}^{\dagger}+a_{0}^{\dagger}a_{1}^{\dagger}b_{0}^{\dagger})\ket{\text{vac}}     \label{eqn:basis_state_sym}, \\
    \ket{\Psi}_{a,a,b}^{\text{mix}}&=\frac{1}{\sqrt{3}}(a_{0}^{\dagger}a_{0}^{\dagger}b_{1}^{\dagger}-a_{0}^{\dagger}a_{1}^{\dagger}b_{0}^{\dagger})\ket{\text{vac}}. 
    \label{eqn:basis_state_mix}
\end{align}
where $a,a,b$ indicates two photons in mode $a$, and one photon in mode $b$, and the indices 0 and 1 refer to the internal states of the photons. Note that the normalisation is absorbed in the creation operators acting on the same mode(s). We now consider the symmetries under the exchange of the two photons in the spatial mode, $a$, with each other, and the second photon in the mode $a$ with the photon in the mode $b$. The state $\ket{\Psi}_{a,a,b}^{\text{sym}}$ is symmetric under the exchange of either pair of photons in their internal and external degrees of freedom. However, the state $\ket{\Psi}_{a,a,b}^{\text{mix}}$ exhibits symmetric behaviour under the exchange of the two photons in the same mode and antisymmetric behaviour when the internal or external states of the second photon in mode $a$, and the photon in mode $b$ are exchanged. The resulting state, therefore, exhibits a so-called \textit{mixed symmetry} (see Table~\ref{table:sym}). 
Note that in all cases, the overall symmetry of the three-photon state always remains symmetric, regardless of the symmetries of the subsystems.

We can now tune between the states $\ket{\Psi}_{a,a,b}^{\text{sym}}$ and $\ket{\Psi}_{a,a,b}^{\text{mix}}$ by introducing a phase parameter $\phi$:
\begin{equation}
    \ket{\Psi^{\phi}}_{a,a,b}= \frac{1}{\sqrt{3}}(a^{\dagger}_{0}a^{\dagger}_{0}b^{\dagger}_{1}+ e^{i\phi} a^{\dagger}_{0}a^{\dagger}_{1}b^{\dagger}_{0})\ket{\text{vac}},\label{eqn:generalised_input_state}
\end{equation}
as depicted in Fig.~\ref{fig:concept}(b).
The state in Eq.~\ref{eqn:generalised_input_state} can be rewritten as 
\begin{equation}
    \ket{\Psi^{\phi}}_{a,a,b}= \cos(\phi/2)\ket{\Psi}_{a,a,b}^{\text{sym}}-i\sin(\phi/2)\ket{\Psi}_{a,a,b}^{\text{mix}},
    \label{eqn:input state v2}
\end{equation}
which reveals that the state is indeed a superposition of the symmetric, $\ket{\Psi}_{a,a,b}^{\text{sym}}$, and the mixed-symmetry state, $\ket{\Psi}_{a,a,b}^{\text{mix}}$, with $\left|\cos(\phi/2)\right|^2=\cos^2(\phi/2)$ and $\left|-i\sin(\phi/2)\right|^2=\sin^2(\phi/2)$ being their respective contributions to the state. The contribution of each component to the scattering statistics is determined by the absolute square of the corresponding amplitude.

The effect of the state's symmetry is revealed through scattering statistics, in particular, by observing the scattering distributions of the photons at a beam splitter. Here we employ a variable beam splitter (VBS), described by the transformation:
\begin{equation}
    U_{\text{VBS}}=
    \begin{pmatrix}
        \cos(\theta)&\sin(\theta)\\
        \sin(\theta)&-\cos(\theta)
    \end{pmatrix},
    \label{eq:VBS}
\end{equation} 
where the parameter $\theta$ describes the splitting ratio.

 The scattering probabilities, $P(n,m)$, denote the probability of detecting $n$ photons in mode $a$ and $m$ photons in mode $b$. Below we give the expressions for the output configurations $(3,0)$ and $(1,2)$:
\begin{align}
 P(3,0)&=\cos^2(\phi/2) \cdot p_{3,0}(\theta), \label{eqn:P30}\\
P(1,2)&=\cos^2(\phi/2) \cdot p_{1,2}(\theta)-\sin^2(\phi/2) \cdot p'_{1,2}(\theta), \label{eqn:P12}
\end{align}
where the quantities $p_{3,0}$, $p_{1,2}$, and $p'_{1,2}$ depend only on the splitting ratio of the VBS $\theta$ (refer to the Appendix for detailed calculations and definitions). As before in Eq.~\ref{eqn:input state v2}, the factors $\cos^2(\phi/2)$ and $\sin^2(\phi/2)$ are the contributions of the symmetric and mixed-symmetry components of the input state, respectively. By isolating $\theta$- and $\phi$-dependence in the expressions, we make explicit the contributions of the beam splitter scattering and the intrinsic symmetry of the state to the scattering statistics.

We see that the fully-bunched output $(3,0)$ only results from the symmetric component of the state, and vanishes when $\phi=\pi$. The partially bunched distributions 
have contributions from the symmetric and mixed-symmetry components of the state and never vanish. The complete set of probabilities, including analogous cases, can be found in the Appendix.

We aim at studying three-photon interference and the influence of symmetry on the scattering statistics. Note that the state in Eq.~\ref{eqn:generalised_input_state} can be constructed by combining a single photon 
with a $\ket{\Psi}$-type Bell state with a generalised phase. That is,
\begin{equation}
    \ket{\Psi^{\phi}}'_{a,a,b}=a^{\dagger}_{0}\cdot \frac{1}{\sqrt{2}}(a^{\dagger}_{0}b^{\dagger}_{1}+ e^{i\phi} a^{\dagger}_{1}b^{\dagger}_{0})\ket{\text{vac}}.
    \label{eqn:Singlephoton_plus_Bell_state_phi}
\end{equation}

\begin{figure*}[t]
    \centering
    \includegraphics[width=1.0\textwidth]{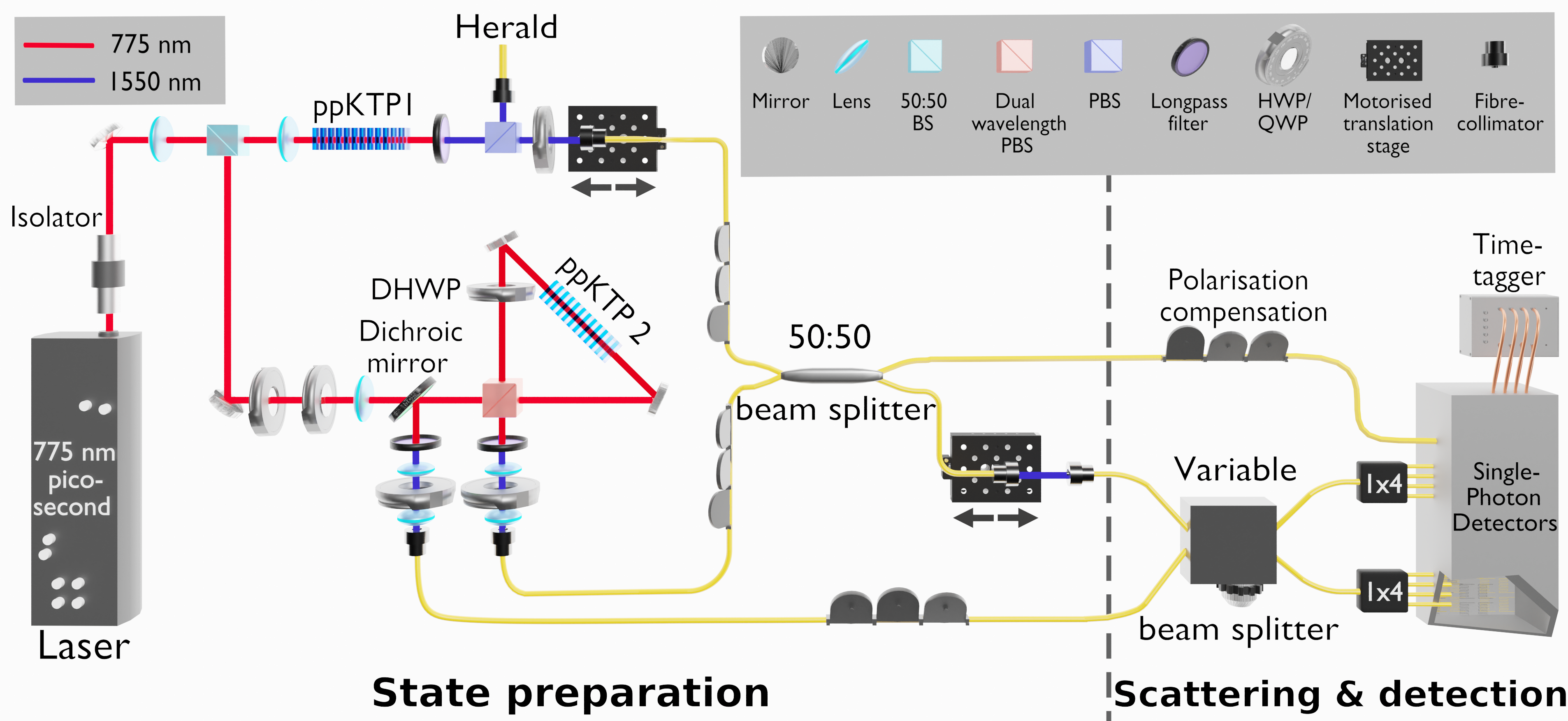}
    \caption{\textbf{Overview of experimental setup:} A pico-second pulsed laser at $\lambda_{p}=775~\textrm{nm}$ is used to pump two spontaneous parametric down-conversion sources consisting of periodically poled potassium titanyl phosphate (ppKTP 1, ppKTP 2) crystals. The resulting down-converted signal and idler photons are at $\lambda_{s/i}=1550~\textrm{nm}$. The first source (ppKTP 1) is a linear source, gives us the heralded single photon, while the other source (ppKTP 2) is a Sagnac-type source, which results in the generation of a Bell state. The heralded single photon and one half of the Bell state are sent to the two input ports of a fibre-based 50:50 beam splitter. One of the outputs of the 50:50 beam splitter is directed to one input to the VBS.  Only the cases where both photons are directed to the VBS and no photons are detected at the other output are considered to be successful measurement runs. The other half of the Bell state is directed to the other input of the VBS. Each output of the VBS is demultiplexed to four superconducting nano-wire single photon detector (SNSPD) channels to have pseudo photon number resolution.}
    \label{fig:experimental_setup}
\end{figure*}
Therefore, we compare the scattering statistics of the three-photon state to those of a Bell state and a single photon that do not interfere. We calculate the scattering statistics of the Bell state and the single photon separately and obtain the product of the statistics $\xi(n,m)$ by multiplying them. The single photon scatters probabilistically with 
\begin{align}
    P_{\text{s}}(1,0)&=\cos^2(\theta),\\
    P_{\text{s}}(0,1)&=\sin^2(\theta),
\end{align}
depending only on the setting of the VBS.

A $\ket{\Psi}$-type Bell state with a general phase, $\phi$,is defined as:
\begin{equation}
    \ket{\Psi^{\phi}}=\frac{1}{\sqrt{2}}(a^{\dagger}_{0}b^{\dagger}_{1}+e^{i\phi}a^{\dagger}_{1}b^{\dagger}_{0})\ket{\text{vac}},
\end{equation}
which can be written as a superposition of the symmetric and anti-symmetric Bell states as 
\begin{equation}
\ket{\Psi^{\phi}}=\cos(\phi/2)\ket{\Psi^+}-i\sin(\phi/2)\ket{\Psi^-},
\end{equation}
where the phase $\phi$ sets the symmetry of the state. 
The output statistics for the scattering of such a state at a VBS are:
\begin{align}
    P_{\text{Bell}}(2,0)&=\frac{1}{2}\cos^2(\phi/2) \sin^2(2\theta)=P_{\text{Bell}}(0,2),\\
    P_{\text{Bell}}(1,1)&=\cos^2(\phi/2)\cos^2(2\theta)-\sin^2(\phi/2)^.
    \label{eq:scatteringprobab_bellstates}
\end{align}
For $\phi=\pi$, we obtain the anti-symmetric Bell state, and the bunching probabilities $P_{\text{Bell}}(2,0)$ and $P_{\text{Bell}}(0,2)$ vanish, while $P_{\text{Bell}}(1,1)$ remains constant. The anti-symmetry of the state prevents the two photons from occupying the same mode. For ${\phi=0}$, we obtain the symmetric Bell state, $\ket{\Psi^+}$, and the scattering statistics depend on the splitting ratio of the VBS (refer to the Appendix).

Now we can calculate the three-photon statistics (i.e., product of statistics), $\xi(n,m)$, from the statistics of the single photon and the Bell state. Here we present the probabilities for $\xi(3,0)$ and $\xi(1,2)$.

\begin{align}
    \begin{split} \label{eqn:xi30}
    \xi(3,0)&=P_{\text{s}}(1,0) \cdot P_{\text{Bell}}(2,0)\\
						&=\cos^2(\phi/2)  \cdot \tilde{p}_{3,0}(\theta),
    \end{split}\\
    \begin{split}
        \xi(1,2)&=P_{\text{s}}(1,0) \cdot P_{\text{Bell}}(0,2)+P_{\text{s}}(0,1) \cdot P_{\text{Bell}}(1,1) \\
    &=\cos^2(\phi/2)\cdot \tilde{p}_{1,2}(\theta)-\sin^2(\phi/2) \cdot \tilde{p}'_{1,2}(\theta),\label{eqn:xi12}
    \end{split}
\end{align}

where, as before, $n(m)$ denotes photons in mode 1(2). We see that the product of the statistics can be written in the same form as Eqs.~\ref{eqn:P30} and \ref{eqn:P12}, however, with different quantities: $\tilde{p}_{3,0}$, $\tilde{p}_{1,2}$, and $\tilde{p}'_{1,2}$. Refer to the Appendix for the complete probability distribution, including the cases $\xi(0,3)$ and $\xi(2,1)$, which have similar expressions.
Only in the antisymmetric case do we find that the product of the statistics is the same as the joint statistics for the partially bunched distributions $(2,1)$ and $(1,2)$ and therefore exhibit identical behaviour. That is,
\[
\left.\begin{aligned}
P(1,2) &= \xi(1,2)\\
P(2,1) &= \xi(2,1)
\end{aligned}\right\}\qquad\text{for }\phi=\pi.
\]
For all other cases, the scattering statistics differ and thus we observe three-photon interference signatures that are different from the product of the statistics of a Bell state along with a single photon. 

\begin{figure*}[bh!]
    \centering
    \vspace{-0.5cm}
    \includegraphics[width=0.95\linewidth]{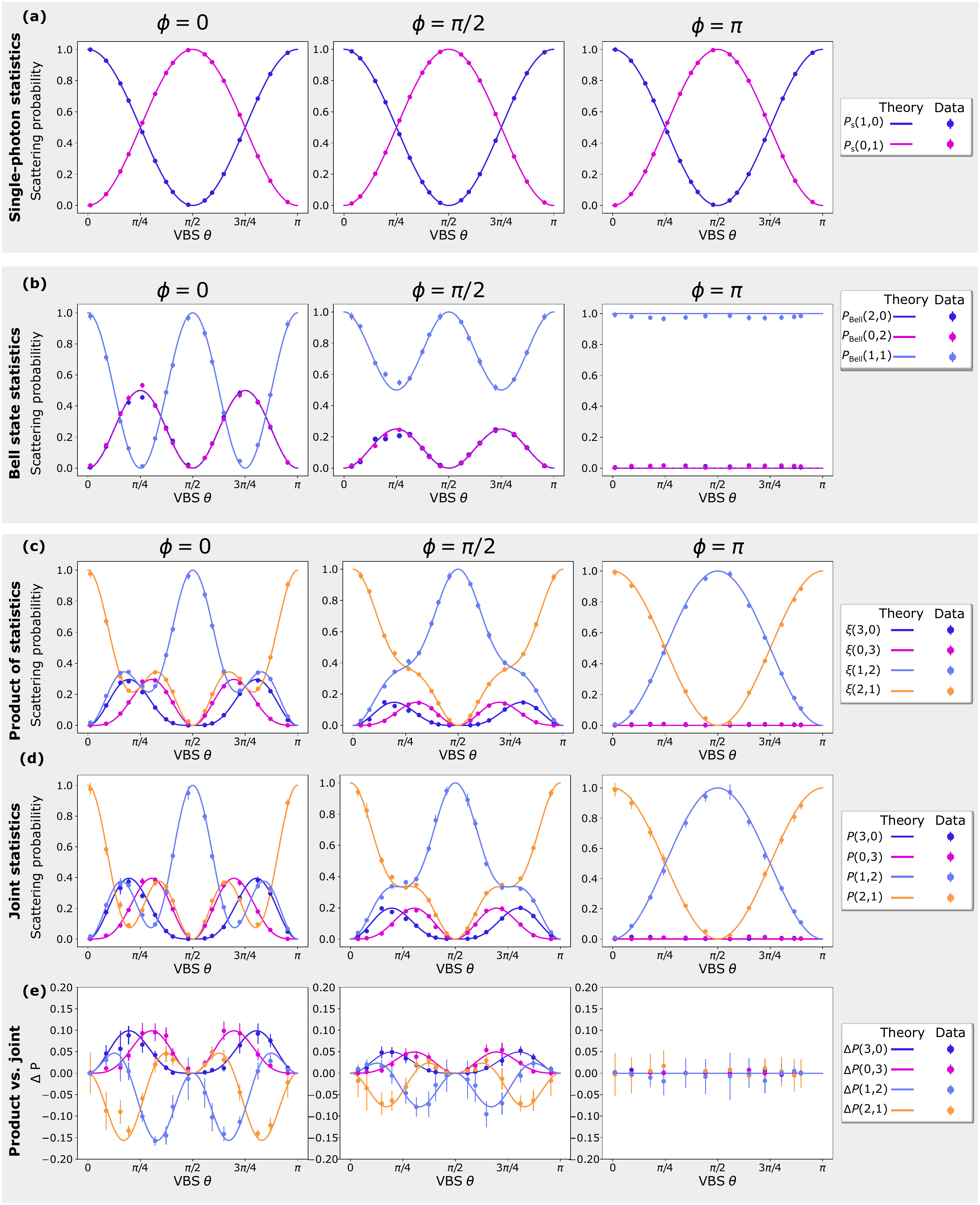}
    \caption{\textbf{Scattering probabilities at the variable beam splitter (VBS):}
    $P(x,y)$ denotes the probability of detecting $x$ photons in the first output mode and $y$ photons in the second. The VBS can be tuned from full transmission ($\theta=0 \text{ and } \theta=\pi$) over a balanced beam splitter ($\theta=\pi/4$ and $\theta=3\pi/4$) to full reflection ($\theta=\pi/2$).
    (a) Scattering statistics for a heralded single photon.
    (b) Scattering statistics for a generalized $\ket{\Psi}$-type Bell state, $\ket{\Psi}_{a,b}^{\phi} = \frac{1}{\sqrt{2}}(a^{\dagger}_{0}b^{\dagger}_{1} + e^{i\phi}a^{\dagger}_{1}b^{\dagger}_{0})\ket{\text{vac}}$, for phases $\phi = \pi$, $\pi/2$, and $0$.
    (c) The product of the independent scattering statistics from panels (a) and (b), denoted $\xi(x,y)$, shown for each value of $\phi$.
    (d) Joint scattering statistics for the three-photon entangled state $\ket{\Psi^{\phi}}_{a,a,b} = \frac{1}{\sqrt{3}}(a^{\dagger}_{0}a^{\dagger}_{0}b^{\dagger}_{1} + e^{i\phi}a^{\dagger}_{0}a^{\dagger}_{1}b^{\dagger}_{0})\ket{\text{vac}}$, also for $\phi = \pi$, $\pi/2$, and $0$.
    (e) The difference $\Delta P = \xi(x,y) - P(x,y)$, highlighting minimal deviation for $\phi = \pi$ and maximal deviation for $\phi = 0$.
    Error bars indicate Poissonian uncertainties arising from photon-counting statistics.}
       \label{fig:Results}
\end{figure*}

\section{Experiment and Results}
We aim to study the interference behaviour of symmetric and mixed-symmetry states experimentally. In our experiment, the internal states are encoded in polarisation with $\ket{0}\equiv\ket{H}$ and $\ket{1}\equiv\ket{V}$. We prepare the three-photon input state by adding a single photon to one of the modes of a generalised Bell state (see Fig.~\ref{fig:concept}(b)). 

The Bell state is generated using a photon source based on spontaneous parametric down conversion (SPDC) in a Sagnac interferometer, and the single photon is generated from a heralded linear SPDC source. The phase of the Bell state can be tuned using a combination of wave plates (see Fig.~\ref{fig:experimental_setup} and the Appendix
for more details). The single photon is combined with one of the photons of the Bell state by using a balanced beam splitter and post-selecting cases where both photons end up in the same spatial mode, leading to the VBS. This results in the desired three-photon state of Eq.~\ref{eqn:generalised_input_state}. The VBS is then employed to tune the coupling between the two modes, and the photon statistics are measured at the output. Each mode of the VBS is demultiplexed to four superconducting nanowire single-photon detectors (SNSPDs) using multiport splitters to achieve pseudo photon number resolution.

Fig.~\ref{fig:Results} presents the measured three-photon scattering statistics, which we use to identify signatures of three-photon interference. Our approach is to compare the joint statistics of the three-photon state with the product of the individual scattering statistics of a two-photon Bell state along with that of a single photon. Deviations between the two statistics indicate signatures of three-photon interference.
We first measure the scattering statistics of the Bell state and the single photon, separately (Fig.~\ref{fig:Results}(a) and (b)) and compute the product of the individual statistics $\xi(n,m)$ (Fig.~\ref{fig:Results}(c)). We then measure the scattering statistics of the three-photon state $P(n,m)$ for different settings of the VBS (Fig.~\ref{fig:Results}(d)).\\

Plotting the difference between these two sets of statistics (Fig.~\ref{fig:Results}(e)) provides a direct visualisation of evident ~\emph{three-photon interference}. This effect can be clearly understood by examining the three cases of $\phi$ separately. 

For $\phi=0$, we have the state $\ket{\Psi}_{a,a,b}^{\text{sym}}$, where we observe a clear difference between the product of the statistics of the Bell state and the single photon, and the measured three-photon statistics. The pronounced deviations indicate strong three-photon interference, demonstrating that the symmetry of the state can enhance multi-photon correlations.

As for $\phi=\pi/2$, The behaviour is intermediate: three-photon interference is still present, but deviations from the product of statistics of the individual components are less pronounced than for {$\phi=0$}, reflecting that the state possessing intermediate symmetry exhibits reduced three-photon interference signatures.

For $\phi=\pi$, as discussed before, the three-photon state $\ket{\Psi}^{\text{mix}}_{a,a,b}$ contains symmetric and anti-symmetric pairs. The anti-symmetric pair always ends up anti-bunching, resulting in the vanishing of the scattering probabilities $P(3,0)$ and $P(0,3)$. Comparing our result with the product of the individual two-photon and single-photon statistics shows that the behaviour is identical in both cases, indicating the absence of genuine three-photon interference present in the three-photon statistics.

We see close agreement between experiment and theory across all data sets, with well-resolved details in the measured scattering statistics. This agreement attests to the high photon quality and precise control of the quantum state achieved in our setup. The small deviations of the measured data from theoretical predictions can be attributed to imperfect Bell state generation and errors that accumulate through the experimental setup. Imperfections in the photon sources, such as slight spectral mismatch, could introduce distinguishability. To check this, we performed Hong-Ou-Mandel interference using the VBS and estimated a visibility of $99.3_{-0.6}^{+0.5}\%$ for the interference of photons generated at the Sagnac source, and $95.4_{-4.5}^{+3.1}\%$ for photons generated from two different sources (refer to the Appendix for more details on the experimental setup). The reduced visibility in the two-source case is due to photons from the two independent sources being slightly different from one another independent photon generation processes. Furthermore, higher-order photon pair generation could result in accidental coincidences that reduce the quality of interference. It was also observed that the VBS exhibited some hysteresis and that it took some time for the splitting ratio to stabilise after being set. Small drifts during the course of the measurements may have contributed to some discrepancies. Combined with Poissonian errors at the detection stage, these effects account for the residual differences between the experimental results and the predicted theory. 

\section{Conclusion}
In this work, we experimentally demonstrated the influence of the symmetry of the constituent states on multiphoton (multiparticle) interference. Using three photons, we were able to access a complex symmetry space that is not possible with smaller systems---one with \textit{mixed-symmetry}, where one pair of photons is symmetric under exchange, while the other is anti-symmetric. By tuning the phase $\phi$, which acts as a direct control parameter over the state's symmetry, we were able to switch on and off genuine three-photon interference. This was evident from the absence of three-photon interference signatures when $\phi$ was set to $\pi$ and the presence of clear deviations between the joint and product of the statistics for $\phi=0$.  

This ability to control the interference by tuning the symmetry of photonic states suggests a scalable approach to probing and emulating particle statistics beyond the bosonic regime. With just two photons, we can already emulate non-bosonic statistics by observing the evolution of their external state. Furthermore, scaling up to three photons allows us to access generalised immanonic statistics that lie between bosonic permanents and fermionic determinants. Extensions to larger multiphoton states, potentially qutrits, could facilitate the emulation of even more complex quantum systems governed by non-trivial exchange statistics. 

Our experiment therefore proves to be a promising platform for exploring the fundamental role of symmetry in quantum interference. With the advent of more advanced hardware, the preparation of different types of photonic entangled states will be increasingly versatile, making the investigated symmetries of wavefunctions ever more accessible.  

\section{Acknowledgments}
We thank Nico Hauser for his inputs on building the Sagnac source, and to Simone Evaldo D’Aurelio for providing the script used to measure photon-number-resolved statistics. We are grateful towards Joscha Heinze for proof-reading the manuscript. We acknowledge the support from the Carl Zeiss Foundation, the Centre for Integrated Quantum Science and Technology (IQST), the Federal Ministry of Research, Technology and Space (BMFTR, projects SiSiQ: FKZ 13N14920, PhotonQ: FKZ 13N15758, QRN: FKZ16KIS2207), the Ministry of Science, Research and Arts Baden-Württemberg (MWK), and the Deutsche Forschungsgemeinschaft (DFG, German Research Foundation, 431314977/GRK2642, SFB 1667). D.B. was supported by the JST Moonshot R\&D program under Grant JPMJMS226C.

\bibliography{refs.bib}
\onecolumngrid
\appendix

\section{Scattering of Bell states}
\label{Appx.A Scattering of Bell States}
The four Bell states $\ket{\Psi^{\pm}}=\frac{1}{\sqrt{2}}(a^{\dagger}_{0}b^{\dagger}_{1}\pm a^{\dagger}_{1}b^{\dagger}_{0})\ket{\text{vac}}$, $\ket{\Phi^{\pm}}=\frac{1}{\sqrt{2}}(a^{\dagger}_{0}b^{\dagger}_{0}\pm a^{\dagger}_{1}b^{\dagger}_{1})\ket{\text{vac}}$ evolve through a two-mode balanced beam splitter as
\begin{align}
    \ket{\Psi^+}&\xrightarrow{BS}\frac{1}{\sqrt{2}}(a^{\dagger}_{0}a^{\dagger}_{1}-b^{\dagger}_{1}b^{\dagger}_{0})\ket{\text{vac}},\\
    \ket{\Psi^-}&\xrightarrow{BS}\frac{-1}{\sqrt{2}}(a^{\dagger}_{0}b^{\dagger}_{1}-a^{\dagger}_{1}b^{\dagger}_{0})\ket{\text{vac}},\label{eq:psi-_BS}\\
    \ket{\Phi^{\pm}}&\xrightarrow{BS}\frac{1}{2\sqrt{2}}((a^{\dagger}_{0}a^{\dagger}_{0}-b^{\dagger}_{0}b^{\dagger}_{0})\pm(a^{\dagger}_{1}a^{\dagger}_{1}-b^{\dagger}_{1}b^{\dagger}_{1}))\ket{\text{vac}},
\end{align}
where $a^{\dagger}_{x}$ and $b^{\dagger}_{x}$ are the creation operators in output modes $a$ and $b$ of the beam splitter, and $x \in \{0,1\}$ denoting the internal degrees of freedom.
$\ket{\Psi^-}$ is the only state that anti-bunches at the beam splitter, while the other three result in bunched statistics. 

\section{Constructing a state with mixed-symmetry}
\label{Appx:Constructing_a_state_with_mixed_symmetry}
A mixed-symmetry state of three particles can be constructed either by first symmetrising and then anti-symmetrising, or vice versa, a three particle wavefunction $\ket{\psi_0}=\ket{a_1,b_2,c_3}$, following ~\cite{porter2009}. We first define the particle permutation operator $\hat{P}_{i,j}$, which swaps the state of particles $i$ and $j$. The corresponding symmetrisation operator is given by {$\hat{S}_{i,j}=(\mathbb{I}+\hat{P}_{i,j})$}, and the anti-symmetrisation operator is given by  {$\hat{A}_{i,j}=(\mathbb{I}-\hat{P}_{i,j})$}. To obtain the mixed symmetry state, we first symmetrise the wavefuntion with respect to particles 1 and 2, and antisymmetrise the pair 2 and 3.
That is, 
\begin{equation}
    \ket{\psi_0}'=\frac{1}{2}({a^{\dagger}_{1}b^{\dagger}_{2}c^{\dagger}_{3}}+{b^{\dagger}_{1}a^{\dagger}_{2}c^{\dagger}_{3}}-{a^{\dagger}_{1}c^{\dagger}_{2}b^{\dagger}_{3}}-{b^{\dagger}_{1}c^{\dagger}_{2}a^{\dagger}_{3}})\ket{\text{vac}}.
\end{equation}

We simplify the scenario by restricting ourselves to only two orthogonal internal states $\ket{0}$ and $\ket{1}$, and two external states (here, spatial modes), $a$ and $b$. Under these conditions, the system can be described by the wavefunction
\begin{equation}
    \begin{aligned}
        \ket{\psi_1}&=\mathcal{N}({a^{\dagger}_{0}a^{\dagger}_{0}b^{\dagger}_{1}}+{a^{\dagger}_{0}a^{\dagger}_{0}b^{\dagger}_{1}}-{a^{\dagger}_{1}a^{\dagger}_{0}b^{\dagger}_{0}}-{a^{\dagger}_{0}a^{\dagger}_{1}b^{\dagger}_{0}})\ket{\text{vac}}\\
        &=2\mathcal{N}({a^{\dagger}_{0}a^{\dagger}_{0}b^{\dagger}_{1}}-{a^{\dagger}_{1}a^{\dagger}_{0}b^{\dagger}_{0}})\ket{\text{vac}},
    \end{aligned}
    \label{eqn:desired_state}
\end{equation}
where $\mathcal{N}$ is the re-normalisation factor and $a^{\dagger}_{0/1}$ and $b^{\dagger}_{0/1}$ are the creation operators for the two internal states in modes $a$ and $b$, respectively. 

\section{Exchange of photons in modes $a$ and $b$}
\begin{table*}[ht!]
    \centering
    \includegraphics[width=0.925\linewidth]{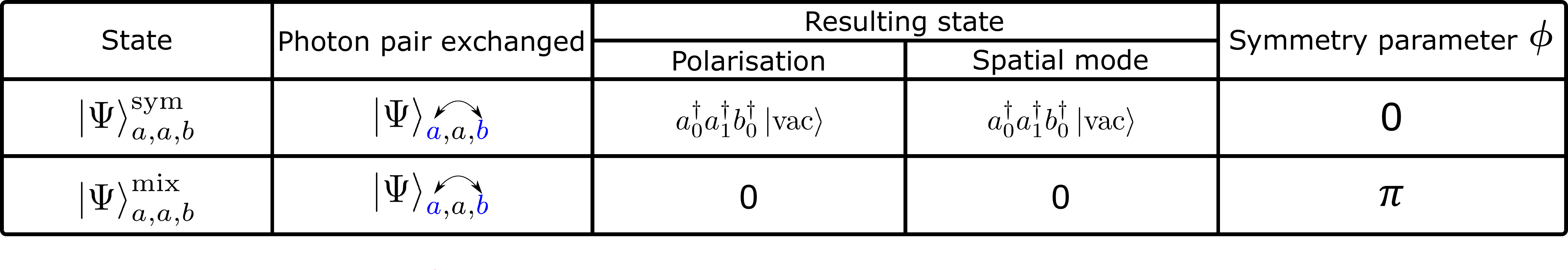}
    \caption{Exchanging different degrees of freedom of photons in modes $a$ and $b$ for the symmetric state $\ket{\Psi}_{a,a,b}^{\text{sym}} = \frac{1}{\sqrt{3}}(a_{0}^{\dagger}a_{0}^{\dagger}b_{1}^{\dagger}+a_{0}^{\dagger}a_{1}^{\dagger}b_{0}^{\dagger})\ket{\text{vac}}$, and the mixed-symmetry state $\ket{\Psi}_{a,a,b}^{\text{mix}}=\frac{1}{\sqrt{3}}(a_{0}^{\dagger}a_{0}^{\dagger}b_{1}^{\dagger}-a_{0}^{\dagger}a_{1}^{\dagger}b_{0}^{\dagger})\ket{\text{vac}}$.}
    \label{fig:exchanging_a_b}
\end{table*}
For completeness, we analyse the exchange of internal or external modes of the photon associated with the state $a^{\dagger}_0\ket{\text{vac}}$ with those of the photon in mode $b$, a case that is not addressed in the main text. 
In the case of the state $\ket*{\Psi}_{a,a,b}^{\text{sym}}$, the symmetric component of the state survives and the resulting state is a product state, whereas for $\ket*{\Psi}_{a,a,b}^{\text{mix}}$, the antisymmetric component cannot exist in a single mode and the state therefore vanishes (see Table~\ref{fig:exchanging_a_b}).

\section{Generating the three-photon state}
\label{InputatVBS}
The state in Eq.~\ref{eqn:desired_state} can be obtained by having a $\ket{\Psi}$-type Bell state, $\ket{\Psi}_{a,b}^{\phi} = \frac{1}{\sqrt{2}}\left(a^{\dagger}_{0}b^{\dagger}_{1} + e^{i\phi}a^{\dagger}_{1}b^{\dagger}_{0}\right)\ket{\text{vac}}$, across two spatial modes $a$ and $b$, along with a single photon in a third spatial mode, $\gamma$, in the state $\gamma^{\dagger}_{0}\ket{\text{vac}}$, which is added to one of the modes of the Bell state, effectively resulting in an additional $a^{\dagger}_{0}\ket{\text{vac}}$. In order to experimentally realise this, we start with a two-photon Bell state across two separate spatial modes, and a single heralded photon in a third spatial mode. The initial state is given by 
\begin{equation}
    \ket{\Psi_{in}}= \frac{1}{\sqrt{2}}(\gamma^{\dagger}_{0}a^{\dagger}_{0}b^{\dagger}_{1}+ e^{i\phi} \gamma^{\dagger}_{0}a^{\dagger}_{1}b^{\dagger}_{0})\ket{\text{vac}},
\end{equation}
where the three modes are denoted by $\gamma$, $a$, and $b$. 
\begin{figure}[b!]
    \centering
    \includegraphics[width=0.45\linewidth]{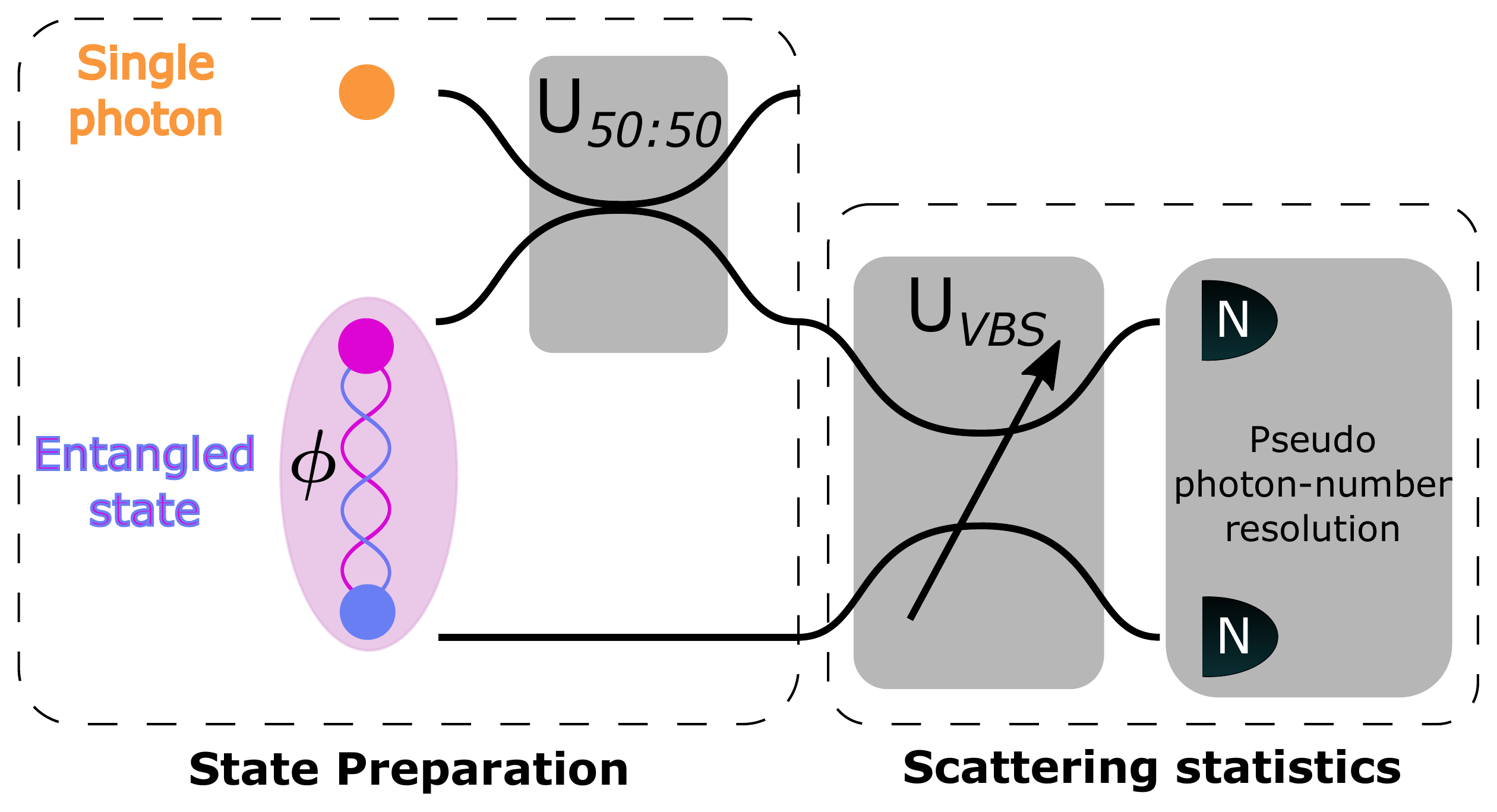}
    \caption{\textbf{Schematic of the experimental setup:} A single photon is injected into the first mode of an entangled state via a 50:50 beam splitter. The desired three-photon state is post-selected by selecting events in which the injected photon and the first photon of the entangled pair occupy the same spatial mode, forming the variable beam splitter (VBS). The splitting ratio is controlled at the VBS, and the resulting scattering statistics are measured using pseudo–photon-number resolution, achieved by demultiplexing each VBS output into four detector channels.}
    \label{fig:}
\end{figure}
The single photon and one photon of the Bell state impinge on a balanced beam splitter, after which we post-select on events where both photons exit the beam splitter at the same output port, particularly the port leading to the variable beam splitter.

The state after the beam splitter is given by 
\begin{equation}
    \ket{\Psi_{in}'}=\frac{1}{\sqrt2}(\frac{1}{2}(\gamma^{\dagger2}_{0}-a^{\dagger2}_{0})b^{\dagger}_{1}+e^{i\phi}\frac{1}{2}(\gamma^{\dagger}_{0}\gamma^{\dagger}_{1}-\gamma^{\dagger}_{0}a^{\dagger}_{1}+\gamma^{\dagger}_{1}a^{\dagger}_{0}-a^{\dagger}_{0}a^{\dagger}_{1})b^{\dagger}_{0})\ket{\text{vac}}.
\end{equation}
The post-selection condition involves discarding all terms containing at least one photon in mode $\gamma$, resulting in the state:
\begin{equation}
    \ket{\Psi_{in}''}=\frac{1}{2\sqrt2}((a^{\dagger2}_{0})b^{\dagger}_{1}+e^{i\phi}(a^{\dagger}_{0}a^{\dagger}_{1})b^{\dagger}_{0})\ket{\text{vac}},
\end{equation}
which, upon renormalisation, results in the desired input state. 

In other words, the single photon and one photon from the Bell state impinge on a beam splitter, and the interference results in the addition of the single photon to the Bell state mode with probability of $P_{\text{addition}}=3/8$.

\section{Output of VBS and scattering probabilities}
\label{output_joint_scattering}
In this section, the transformation of the three-photon input state in Eq.~\ref{eqn:generalised_input_state} through a lossless VBS is derived. The input state is given by 
\begin{equation}
    \ket{\Psi^{\phi}}_{a,a,b}= \frac{1}{\sqrt{3}}(a^{\dagger}_{0}a^{\dagger}_{0}b^{\dagger}_{1}+ e^{i\phi} a^{\dagger}_{0}a^{\dagger}_{1}b^{\dagger}_{0})\ket{\text{vac}},
   \end{equation}
which can be rewritten as 
\begin{align}
     \ket{\Psi^{\phi}}_{a,a,b}&=\frac{1}{\sqrt{3}}e^{i\phi/2}(e^{-i\phi/2}a^{\dagger}_{0}a^{\dagger}_{0}b^{\dagger}_{1}+ e^{i\phi/2} a^{\dagger}_{0}a^{\dagger}_{1}b^{\dagger}_{0})\ket{\text{vac}}\\
    &\equiv\frac{1}{\sqrt{3}}(\cos{(-\phi/2)}a^{\dagger}_{0}(a^{\dagger}_{0}b^{\dagger}_{1}+a^{\dagger}_{1}b^{\dagger}_{0})+ i\sin(-\phi/2) a^{\dagger}_{0}(a^{\dagger}_{0}b^{\dagger}_{1}-a^{\dagger}_{1}b^{\dagger}_{0}))\ket{\text{vac}}\\
    &=\frac{1}{\sqrt{3}}(\underbrace{\cos{(-\phi/2)}(a^{\dagger}_{0}a^{\dagger}_{0}b^{\dagger}_{1}+a^{\dagger}_{0}a^{\dagger}_{1}b^{\dagger}_{0})}_{\text{symmetric}}+\underbrace{i\sin(-\phi/2)(a^{\dagger}_{0}a^{\dagger}_{0}b^{\dagger}_{1}-a^{\dagger}_{0}a^{\dagger}_{1}b^{\dagger}_{0})}_{\text{mixed-symmetry}})\ket{\text{vac}}.
\end{align}
This representation of the input state makes the distinction between the symmetric and mixed-symmetry components explicit.

The VBS is ideally polarisation insensitive and transforms the input operators $a^{\dagger}_{X}\rightarrow \cos{(\theta)}a^{\dagger}_{X}+\sin(\theta) b^{\dagger}_{X}$, and $b^{\dagger}_{X}\rightarrow \sin{(\theta)}a^{\dagger}_{X}-\cos(\theta) b^{\dagger}_{X}$. The state obtained after the evolution of the three-photon state through the VBS is therefore,
\begin{equation}
\begin{aligned}
\ket{\Psi^{\phi}}_{a,a,b}^{VBS}=& \frac{1}{\sqrt{3}}(\cos(-\phi/2)(\cos(\theta) a^{\dagger}_{0}+\sin(\theta) b^{\dagger}_{0})\cdot(\sin(2\theta)\cdot(a^{\dagger}_{0}a^{\dagger}_{1}-b^{\dagger}_{0}b^{\dagger}_{1})-\cos(2\theta)\cdot(a^{\dagger}_{0}b^{\dagger}_{1}+a^{\dagger}_{1}b^{\dagger}_{0}))\\
    &+i\sin(-\phi/2)(\cos(\theta) a^{\dagger}_{0}+\sin(\theta)b^{\dagger}_{0})\cdot(a^{\dagger}_{1}b^{\dagger}_{0}-a^{\dagger}_{0}b^{\dagger}_{1}))\ket{\text{vac}}.
\end{aligned}
\end{equation}
The above expression can be expanded, and the terms can be grouped according to the photon number distributions:
\begin{align}
    \ket{\Psi^{\phi}}_{a,a,b}^{VBS}=&\frac{1}{\sqrt{3}}(\underbrace{(\cos(-\phi/2)\cos(\theta)\sin(2\theta)) a^{\dagger}_{0}a^{\dagger}_{0}a^{\dagger}_{1}}_{\text{(3,0)}}-\underbrace{(\cos(-\phi/2)\sin(\theta)\sin(2\theta))b^{\dagger}_{0}b^{\dagger}_{0}b^{\dagger}_{1}}_{\text{(0,3)}} \\
    &-\underbrace{(\cos(-\phi/2)\cos(\theta)\sin(2\theta)+\cos(-\phi/2)\sin(\theta)\cos(2\theta)+i\sin(-\phi/2)\sin(\theta))a^{\dagger}_{0}b^{\dagger}_{0}b^{\dagger}_{1}}_{\text{(1,2)}}\\
    &-\underbrace{(\cos(-\phi/2)\cos(\theta)\cos(2\theta)+i\sin(-\phi/2)\cos(\theta))a^{\dagger}_{0}a^{\dagger}_{0}b^{\dagger}_{1}}_{\text{(2,1)}} \\
    &+\underbrace{(\cos(-\phi/2)\sin(\theta)\sin(2\theta)-\cos(-\phi/2)\cos(\theta)\cos(2\theta)+i\sin(-\phi/2)\cos(\theta))a^{\dagger}_{0}a^{\dagger}_{1}b^{\dagger}_{0}}_{\text{(2,1)}}\\
    &-\underbrace{(\cos(-\phi/2)\sin(\theta)\cos(2\theta)-i\sin(-\phi/2)\sin(\theta))a^{\dagger}_{1}b^{\dagger}_{0}b^{\dagger}_{0}}_{\text{(1,2)}})\ket{\text{vac}} ,
\end{align}
where the labels indicate the scattering distributions to which the corresponding terms contribute. The scattering probabilities are computed by taking the square modulus of the specific terms. Note that the cases with more than one photon in a specific mode results in a normalisation factor. 
\begin{align}
\begin{split}
    P(3,0)&=\left|\sqrt{\frac{2}{3}}\cos(-\phi/2)\cos(\theta)\sin(2\theta)\right|^2 =\frac{8}{3}|\cos(-\phi/2)|^2\cos^4(\theta)\sin^2(\theta),
\end{split}\\
\begin{split}
    P(0,3)&=\left|\sqrt{\frac{2}{3}}\cos(-\phi/2)\sin(\theta)\sin(2\theta)\right|^2 =\frac{8}{3}\left|\cos(-\phi/2)\right|^2\sin^4(\theta)\cos^2(\theta),
\end{split}\\
\begin{split}
     P(1,2)&=\frac{1}{3}\Bigl(\left|\cos(-\phi/2)\cos(\theta)\sin(2\theta)+\cos(-\phi/2)\sin(\theta)\cos(2\theta)+i\sin(-\phi/2)\sin(\theta)\right|^2 \\
     &\phantom{={}} +\left|\sqrt{2}\cos(-\phi/2)\sin(\theta)\cos(2\theta)+\sqrt{2}i\sin(-\phi/2)\sin(\theta)\right|^2\Bigr),\\
     &=\frac{1}{3}\left|\cos(-\phi/2)\right|^2((\cos(\theta)\sin{(2\theta)}+\sin(\theta)\cos{(2\theta)})^2+2(\sin(\theta)\cos{(2\theta)})^2)+\left|i\sin(-\phi/2)\right|^2\sin^2(\theta),
\end{split}\\
\begin{split}
     P(2,1)&=\frac{1}{3}\Bigl(\left|\cos(-\phi/2)\cos(\theta)\cos(2\theta)-\cos(-\phi/2)\sin(\theta)\sin(2\theta)+i\sin(-\phi/2)\cos(\theta)\right|^2\\
     &\phantom{={}}+\left|-\sqrt{2}\cos(-\phi/2)\cos(\theta)\cos(2\theta)+\sqrt{2}i\sin(-\phi/2)\cos(\theta)\right|^2\Bigr)\\
     &=\frac{1}{3}\left|\cos(-\phi/2)\right|^2((\cos(\theta)\cos(2\theta)-\sin(\theta)\sin2(\theta))^2+2(\cos(\theta)\cos(2\theta))^2)+\left|i\sin(-\phi/2)\right|^2\cos^2(\theta),
\end{split}
\end{align}

From this representation, it is clear that the bunched statistics appear only for the symmetric state (for $\phi=0$), and vanish for the mixed-symmetry state ($\phi=\pi$). This is due to the fact that the symmetric component of the state allows bunching behaviour, whereas the antisymmetric component always anti-bunches, preventing the appearance of the distributions $(3,0)$ and $(0,3)$. However, partially bunched statistics are present for both symmetries. 

Comparing these probabilities with Eqs.~\ref{eqn:P30} and ~\ref{eqn:P12} in the main text, the quantities $p_{x,y}$ can be identified as 
\begin{equation}
\begin{aligned}
p_{3,0}(\theta)&=\frac{8}{3}\cos^4(\theta)\sin^2(\theta),
    \label{eqn:Phi30}\\
p_{1,2}(\theta)&=\frac{1}{3}((\cos(\theta)\sin{(2\theta)}+\sin(\theta)\cos{(2\theta}))^2+2(\sin(\theta)\cos{(2\theta)})^2)\\
p'_{1,2}(\theta)&=\sin^2(\theta).
\end{aligned}
\end{equation}

\section{Calculating the product of statistics}
\label{product_statistics}

In this section, we derive the exact form of the product of statistics $\xi(n,m)$ and the quantities $\tilde{p}_{3,0}$, $\tilde{p}_{1,2}$, and $\tilde{p}'_{1,2}$. We start from 
\begin{align}
    \xi(3,0)&=P_{\text{s}}(1,0) \cdot P_{\text{Bell}}(2,0),\\
    \xi(0,3)&=P_{\text{s}}(0,1) \cdot P_{\text{Bell}}(0,2),\\
    \xi(1,2)&=P_{\text{s}}(1,0) \cdot P_{\text{Bell}}(0,2)+P_{\text{s}}(0,1) \cdot P_{\text{Bell}}(1,1), \\
    \xi(2,1)&=P_{\text{s}}(1,0) \cdot P_{\text{Bell}}(1,1)+P_{\text{s}}(0,1) \cdot P_{\text{Bell}}(2,0).
\end{align}
Inserting the different probabilities derived for the single photon and the Bell state, we find
\begin{equation}
\begin{aligned}
    \xi(3,0)&= \cos^2(\theta) \cdot \frac{1}{2}  \cos^2(\phi/2) \sin^2(2\theta) = \cos^2(\phi/2) \tilde{p}_{3,0}, \\
    \xi(0,3)&= \sin^2(\theta) \cdot \frac{1}{2}  \cos^2(\phi/2) \sin^2(2\theta) =  \cos^2(\phi/2) \tilde{p}_{0,3},\\
    \xi(1,2)&= \cos^2(\theta) \frac{1}{2}  \cos^2(\phi/2) \sin^2(2\theta) + \sin^2(\theta) (\cos^2(\phi/2)\cos^2(2\theta)- \sin(\phi/2)^2) \\
    &= \cos^2(\phi/2) \tilde{p}_{1,2} - \sin(\phi/2)^2 \tilde{p}_{1,2}' ,\\
    \xi(2,1)&= \cos^2(\theta) (\cos^2(\phi/2)\cos^2(2\theta)+ \sin(\phi/2)^2) + \sin^2(\theta) \frac{1}{2}  \cos^2(\phi/2) \sin^2 (2\theta) \\
    &= \cos^2(\phi/2) \tilde{p}_{2,1} - \sin^2(\phi/2) \tilde{p}_{2,1}' .
\end{aligned} 
\end{equation}

Therefore, the quantities in Eqs.~\ref{eqn:xi30} and ~\ref{eqn:xi12} are given by
\begin{equation}
\begin{aligned}
\tilde{p}_{3,0}(\theta)&=\frac{1}{2}\sin^2 (2\theta)\cos^2(\theta),
    \label{eqn:TildePhi30}\\
\tilde{p}_{1,2}(\theta)&= \frac{1}{2} \cos^2(\theta) \sin^2(2\theta) + \sin^2(\theta)\cos^2(2\theta),\\ 
\tilde{p}'_{1,2}(\theta)&=\sin^2(\theta) = {p}'_{1,2}(\theta).\\
\end{aligned}
\end{equation}

\section{Experimental details}\label{appx:setup}
\label{Appx_experimental_details}
This section provides a detailed description of the experimental setup used to measure the scattering statistics of the Bell state along with the single photon. The experimental setup comprises two photon sources based on SPDC, a variable beam splitter, and pseudo photon-number-resolving detection. The complete setup is illustrated in Fig. \ref{fig:experimental_setup} of the main text. 
\begin{figure}[b!]
    \centering
    \includegraphics[width=0.8\linewidth]{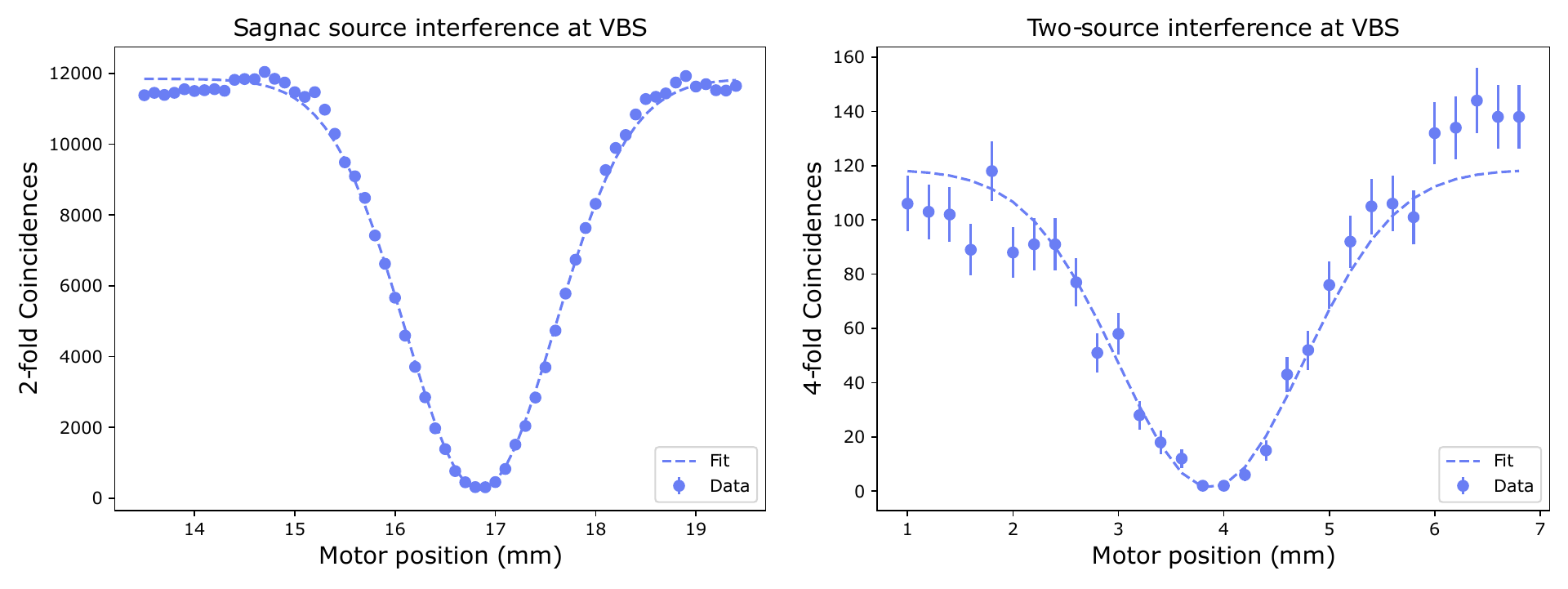}
    \caption{Hong–Ou–Mandel (HOM) interference dips measured for photons from a single source (here, the Sagnac source) and from two independent sources.}
    \label{fig:HOM_dip}
\end{figure}

\textbf{Photon sources:} The photon-pair sources are realised using periodically poled potassium titanyl phosphate (ppKTP) crystals, driven by a pulsed laser at 775 nm. Each crystal is of dimension 1~mm x 1~mm x 30~mm and a poling period of $\Lambda=46.175~\mu\text{m}$. Photon pairs are generated by type-II SPDC, resulting in orthogonally polarised photons at 1550~nm. One of the crystals is placed in a linear configuration and is used to generate a heralded single photon, while the other is placed in a Sagnac-type interferometer, resulting in the generation of a Bell state. An optical isolator is placed at the laser output to suppress counter-propagating beams. In the Sagnac source, a half-wave plate (HWP) sets the pump beam to be diagonally polarised, and in combination with a quarter-wave plate (QWP), the phase of the pump beam and, thereby, the phase of the Bell state can be tuned. A dichroic polarising beam splitter (DPBS) and two mirrors form the Sagnac-loop. The pump beam gets divided at the DPBS, resulting in the crystal being pumped from both the clockwise and counter-clockwise directions. A dual-wavelength HWP (DWHWP) is placed inside the loop to rotate the polarisation of the pump beam in the reflected arm of the DPBS. The bidirectional pumping process produces indistinguishable down-converted photons, resulting in the generation of polarisation-entangled photon pairs, which are coupled into single-mode fibres. The photons travelling co-linearly to the pump beam are separated using a dichroic mirror (DM). Long-pass filters are used to block the pump beam in both sources. In addition, narrow-band 1.5~nm filters centred at 1550~nm are used to ensure spectral indistinguishability. We measured the Hong-Ou-Mandel (HOM) interference visibility of the photons from the Sagnac source as well as the visibility between the heralded single photon and one of the modes of the Bell state by setting the VBS to 50:50 splitting ratio (Fig.~\ref{fig:HOM_dip}). We estimate the visibilities to be $99.5\pm{0.7}\%$ and $98\pm{5}\%$ for the two cases, respectively. We truncated the visibilities and uncertainties by assuming that they follow a Gaussian probability distribution each. We then normalised the cumulative distribution functions in the range from 0 to 100\% and redefined our results as the medians and 68\% credibility intervals of these normalised functions, resulting in a visibility of $99.3_{-0.6}^{+0.5}\%$ for the photons from the Sagnac source and $95.4_{-4.5}^{+3.1}\%$ for photons from two different sources.  
The generated Bell state is of the form:
\begin{equation}
    \ket{\Psi^{\phi}}=\frac{1}{\sqrt{2}}(a^{\dagger}_{0}b^{\dagger}_{1}+ e^{i\phi} a^{\dagger}_{1}b^{\dagger}_{0})\ket{\text{vac}},
    \label{eqn:Bell_state_generated}
\end{equation}
where $\phi$ is the tunable phase parameter.\\

\textbf{Input state preparation:} In order to construct the three photon input state in Eq. \ref{eqn:desired_state}, one photon from the Bell state and the heralded single photon are directed into the two input ports of a 50:50 fiber beam splitter (Thorlabs, TN1550R5A2). One of the outputs of the beam splitter is coupled out at a free-space delay-line before being routed to the input of the variable beam splitter (VBS), ensuring temporal overlap with the second half of the Bell state. The other output port is monitored by a detector, and only events with no counts in this port are retained.

\textbf{Interferometer:} The interferometer comprises a two-port, fibre-based VBS (Newport F-CPL-1550-N-FP), whose splitting ratio is tuned by varying the effective distance between the two fibres. The device is calibrated by sending photons in one of the inputs and monitoring the splitting ratio. Fig.~\ref{fig:Results}(a) in the main text shows the scattering probability of the heralded single photon as a function of $\theta$, the setting of the VBS. 

\textbf{Photon detection:} The output ports of the VBS are demultiplexed to four detector channels each, which are monitored by a superconducting nanowire single-photon detector (SNSPD). This configuration enables pseudo photon-number resolution. The SNSPDs exhibit detection efficiencies exceeding 90$\%$ at 1550~nm. This process is intrinsically probabilistic, and therefore a correction factor needs to be applied to different photon number distributions. The probability that $n$ photons are resolved by $k$ detector channels is given by 
\begin{equation}
    P(n,k)=\frac{k!}{(k-n)!k^n}.
\end{equation}
The measured rate of $n$ photons in $k$ modes is $P(n,k)$ times the actual rate going through the demultiplexed mode. Therefore, a correction factor of $1/P(n,k)$ is applied to each output mode, dependent on the photon number being resolved.
\end{document}